\documentclass[12pt,preprint]{aastex}

\usepackage{natbib}
\begin{document}
\title{Multi-wavelength spectroscopic observation of EUV jet in AR 10960}

\author{Y. Matsui, T. Yokoyama and N. Kitagawa}
\affil{Department of Earth and Planetary Science, University of Tokyo,
7-3-1 Hongo, Bunkyo-ku, Tokyo 113-0033, Japan}
\and
\author{S. Imada}
\affil{Solar-Terrestrial Environment Laboratory, Nagoya University, Furo-cho, Chikusa-ku, Nagoya 464-8601, Japan}
\begin{abstract}
We have studied the relationship between the velocity and temperature of
 a solar EUV jet.
The highly accelerated jet occurred in the active region NOAA 10960 on 2007 June 5.
Multi-wavelength spectral observations with EIS/{\it Hinode} allow us to investigate Doppler velocities at the wide temperature range.
We analyzed the three-dimensional angle of the jet from the stereoscopic
 analysis with {\it STEREO}.
Using this angle and Doppler velocity, we derived the true velocity of
 the jet.
As a result, we found that the cool jet observed with \ion{He}{2} 256 \AA \
 $\log_{10}T_e[\rm{K}] = 4.9$ is accelerated to around $220 \rm{km/s}$ which is
 over the upper limit of the chromospheric evaporation.
The velocities observed with the other lines are under the upper
 limit of the chromospheric evaporation while most of the velocities
 of hot lines are higher than that of cool lines.
We interpret that the chromospheric evaporation and magnetic
 acceleration occur simultaneously.
 A morphological interpretation of this event based on the reconnection
 model is given by utilizing the multi-instrumental observations.

\end{abstract}

\section{Introduction}
Solar jets are known as the plasma ejection in the solar corona and
strongly believed to be produced by the magnetic reconnection.
X-ray jets were first observed
\citep{1992PASJ...44L.173S,1992PASJ...44L.161S} with the Soft X-ray Telescope
(SXT: \citealt{1991SoPh..136...37T}) aboard {\it Yohkoh}
\citep{1991SoPh..136....1O}.
\cite{1999SoPh..190..167A} observed the solar jets in the extreme
ultraviolet (EUV) with {\it Transition Region and Coronal Explorer}
({\it TRACE}: \citealt{1999SoPh..187..229H}), and revealed those detailed
structures. The statistical studies of X-ray jets were presented by
\cite{1996PASJ...48..123S} using statistics from SXT/{\it Yohkoh} and
\cite{2007PASJ...59S.771S} which were obtained by using X-Ray
Telescope (XRT: \citealt{2007SoPh..243...63G}) aboard {\it Hinode}
\citep{2007SoPh..243....3K}.

The physical model of jets was established by
\cite{1994xspy.conf...29S}. In this model, the trigger of the jet is
the magnetic reconnection between the emerged flux and preexisting coronal magnetic field.
There are two types of acceleration mechanisms in this model.
One is the magnetic acceleration by the magnetic reconnection and the other
is the thermal acceleration due to chromospheric evaporation.
Since the jet is caused by the magnetic reconnection, at least a part of
the jet is accelerated by the magnetic force and its velocity reaches
the Alfv\'en velocity.
Two-dimensional MHD simulations performed by \cite{1995Natur.375...42Y,1996PASJ...48..353Y} successfully reproduced
the magnetic acceleration by the magnetic reconnection.
The blowout jet \citep{2010ApJ...720..757M} is a kind of the magnetic
acceleration mechanism.
In this model, the magnetic reconnection causes blowout eruption of the cool
filament.
In the reconnection model, the chromospheric evaporation
(i.e. acceleration by the high pressure caused through the thermal
energy input from the magnetic reconnection site or non-thermal particles) occurs simultaneously.
\cite{2001ApJ...550.1051S} performed one-dimensional hydrodynamic
simulations and succeeded in reproducing the evaporation flow.
 \cite{2004ApJ...614.1042M} succeeded in reproducing the
evaporation flow caused by magnetic reconnection in MHD simulations.
The velocity by the thermal acceleration depends on the temperature.
This temperature dependence was observed in the chromospheric evaporation
by the flare \citep{2009ApJ...699..968M} and dimming flows \citep{2007PASJ...59S.793I,2011ApJ...743...57I}.
A theoretical upper limit of the chromospheric evaporation velocity is
approximately $2.35C_s$, where $C_s$ is the sound speed \citep{1984ApJ...281L..79F}.

In order to analyze the acceleration of the jet, the spectroscopic
observation is effective because the temperature and velocity of the
plasma are obtained from the spectroscopic observation.
However previous spectroscopic observations
\citep{2007PASJ...59S.751C,2010A&A...510L...1K} did not reveal the
mechanism of the acceleration of the jet. 
 Our objective in this study is to establish the detailed acceleration
 mechanism of the solar jets with the multi-wavelength spectroscopic
 observation.
In this paper, we study simultaneous observations by multiple
instruments of a jet on 2007 June 5.
We mainly use the data of EUV imaging
spectrometer (EIS: \citealt{2007SoPh..243...19C}) aboard {\it Hinode} which
is very useful in understanding the plasma dynamics of the jet.
We also use XRT aboard {\it Hinode}, {\it TRACE} 171 \AA \  and 1600 \AA \  data to
study the detailed morphology and temporal variation, Sun Earth Connection
Coronal and Heliospheric Investigation (SECCHI:
\citealt{2008SSRv..136...67H})/EUV Imager (EUVI:
\citealt{2004SPIE.5171..111W}) aboard {\it Solar Terrestrial Relations Observatory} ({\it STEREO}: \citealt{2008SSRv..136....5K})
to study the inclination angle of the jet, Michelson Doppler Imager (MDI:
\citealt{1995SoPh..162..129S}) aboard {\it Solar and Heliospheric Observatory}
({\it SOHO}: \citealt{1995SoPh..162....1D}) to obtain the magnetic
field, and Solar Magnetic Activity Research Telescope (SMART:
\citealt{2004SPIE.5492..958U}) at Hida Observatory to investigate the
response of the chromosphere.

\section{Data Set}
EIS has high spectral and spatial resolution and obtains the imaging and
spectral data simultaneously by the raster scan.
On 2007 Jun 5 04:16 UT, EIS started the raster scan on the active region
NOAA 10960 which consisted of 120 slit positions with the slit width of
2'' and height 240'' and with the field of view of 240'' $\times$ 240''.
Each exposure was 5 s, the cadence was 6 s and the total duration was 12
min.
The data in 17 spectral windows are obtained for this data set.
We calibrate the EIS data using the standard processing routine
eis\_prep in SolarSoftWare, which removes the dark current,
cosmic-ray affected data, and hot pixels.
The slit tilt and the orbital variation of the line position are
corrected using the eis\_tilt\_correction routine and the
eis\_orbit\_spline routine, respectively.

We also use XRT.
All XRT data in this study were taken with the Ti\_poly
filter.
The maximum of the temperature response of this filter is close to 8
MK (\citealt{2004ASPC..325...15K}) and this temperature is higher
than all of the emission lines which we analyze in this paper.
XRT data are calibrated by using xrt\_prep in SolarSoftWare and
normalized with each exposure time.
XRT was observed at the unequal cadence and there were
time when XRT was not observed.
{\it TRACE} has a capability to take EUV images with high spatial
 resolution of 1''.
Thus, its data are useful to study the detail of the EUV jet's structures and
 temporal variation.
 When EIS observed NOAA 10960, {\it TRACE} also observed this active region in
 the \ion{Fe}{9}/\ion{Fe}{10} 171 \AA \ with a cadence of 60 s.
 The {\it TRACE} field of view is 8.5' $\times$ 8.5', and the EIS field of view is
 covered by {\it TRACE} in this observation.
 {\it TRACE} also observed this active region with 1600 \AA \  with a cadence
 of almost one hour.
 We use one 1600 \AA \ image at 04:31 UT.
 {\it TRACE} data are corrected by standard processing routine
trace\_prep, which removes the dark current,
cosmic rays, and a readout pedestal.

EUVI data from both
{\it STEREO-A} and {\it STEREO-B} are also available.
We use EUVI to calculate an inclination angle of the jet.
The separation angle in terms of the sun between {\it STEREO-A} and
{\it STEREO-B} was $11^\circ$ when {\it STEREO-A} and {\it STEREO-B} observed the jet, which is suitable to calculate the inclination angle of the jet.
 Solar Optical Telescope aboard {\it Hinode} (SOT:
 \citealt{2008SoPh..249..167T}) did not observe this active region, so we
 cannot get the detailed structure of the magnetic field.
 Instead, MDI full-disk magnetograms are obtained. The cadence was 96 min,
 and we use the data at 03:15 UT.
The H$\alpha$ images of this active region are available by SMART-T3 around 2007 Jun 5 04:16 UT.

All images are co-aligned with each other. The co-alignment of the images
from different instruments is done based on structures seen at similar temperatures.
First we compare the MDI magnetogram with {\it TRACE} 1600 \AA \ image and
co-align them with the position of the sunspot.
{\it TRACE} 171 \AA \  and 1600 \AA \  images are co-aligned at 04:31UT, then 
other 171 \AA \  images are de-rotated to a reference time of 04:31 UT.
EIS 195 \AA \  images are compared with {\it TRACE} 171 \AA \  at 04:18UT.
The relative positional offset in the data sets of EIS obtained by its
two CCDs (in bands 246-292 \AA \  and 170-211 \AA) is corrected by using
the procedure eis\_ccd\_offset in SolarSoftWare.
XRT images are co-aligned with {\it TRACE} 171 \AA \ images.
Finally, we compare SMART-T3 H$\alpha$ image with the MDI magnetogram
 and co-aligned them using the position of the sunspot.

\section{Results}
\subsection{Three-dimensional Structure of the Jet\label{angle}}
\begin{figure}
\begin{center}
 \includegraphics[width=100mm]{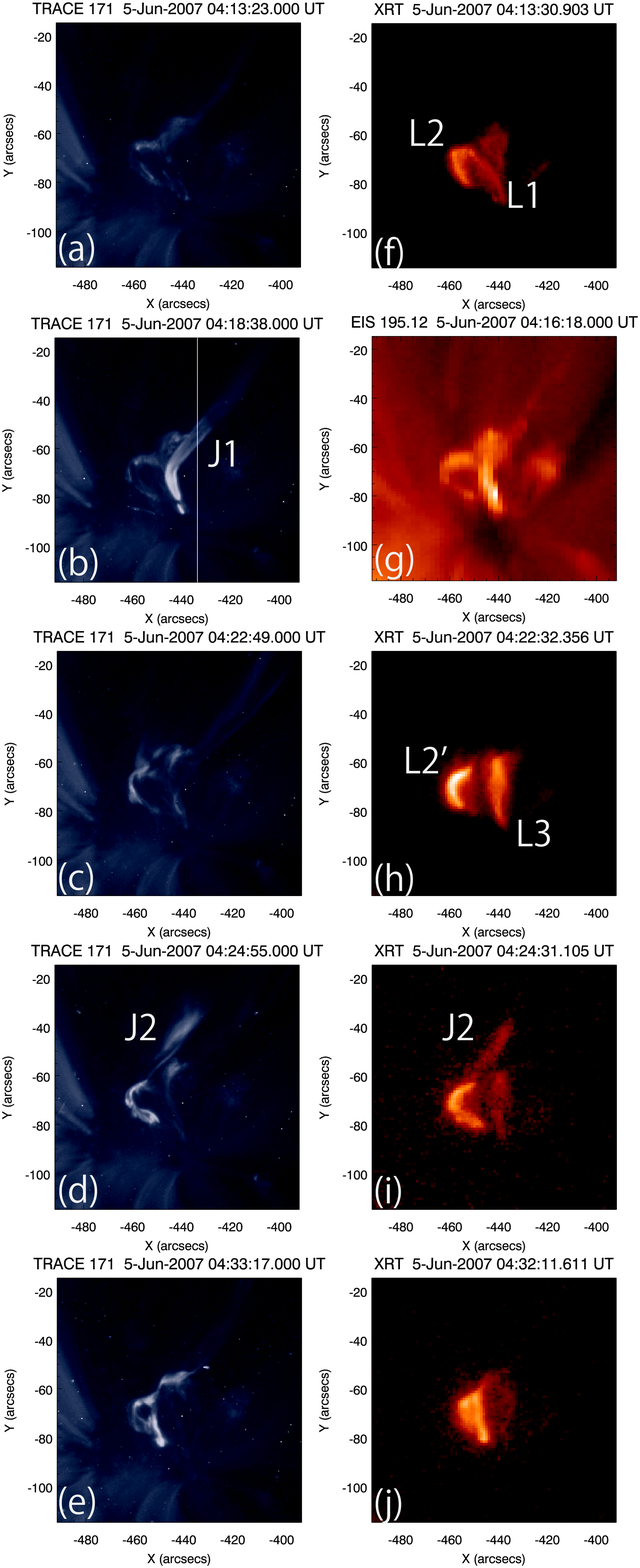}
\caption{(a)-(e) are EUV images observed with {\it TRACE} 171 \AA.
  The vertical white line in (b) corresponds to the position
 of the EIS slit at the {\it TRACE} observation time.
 (f), (h), (i) and (j) are X-ray images observed with XRT with the
 Ti\_poly filter.
 (g) is the EIS intensity image in \ion{Fe}{12} 195 \AA.
 \label{fig1}}
\end{center}
\end{figure}

The jet occurred in the active region NOAA 10960 (S03, E25)
around 04:18 UT, 2007 June 5.
Figure \ref{fig1} shows the images observed with {\it TRACE},
XRT and EIS. Figures 1 (a)-(e) show EUV images observed
with {\it TRACE} 171 \AA.
An EUV jet J1 is shown in figure \ref{fig1}(b). After J1 was observed, a
second EUV jet J2 was observed (figure \ref{fig1}(d)).
Figures \ref{fig1}(f), (h), (i), and (j) show
the X-ray images observed with XRT Ti\_poly filter. Figure \ref{fig1}(f)
shows bright loops L1 and L2 which appear before J1.
The X-ray image of J1 is unavailable due to the data gap of the XRT
observation.
After J2, there are new loops (figure \ref{fig1}(h)).
After J1 occurred, L2 changes into L2' and L3 appears.
Figure \ref{fig1}(i) shows J2 in X-ray at the same time of figure \ref{fig1}(d).
EIS observed J1 with a raster scan.
Figure \ref{fig1}(g) shows the EIS intensity map in \ion{Fe}{12} 195 \AA.
In this paper, we analyze the multi-wavelength spectroscopic observation
of J1.

\begin{figure}
  \includegraphics[width=170mm]{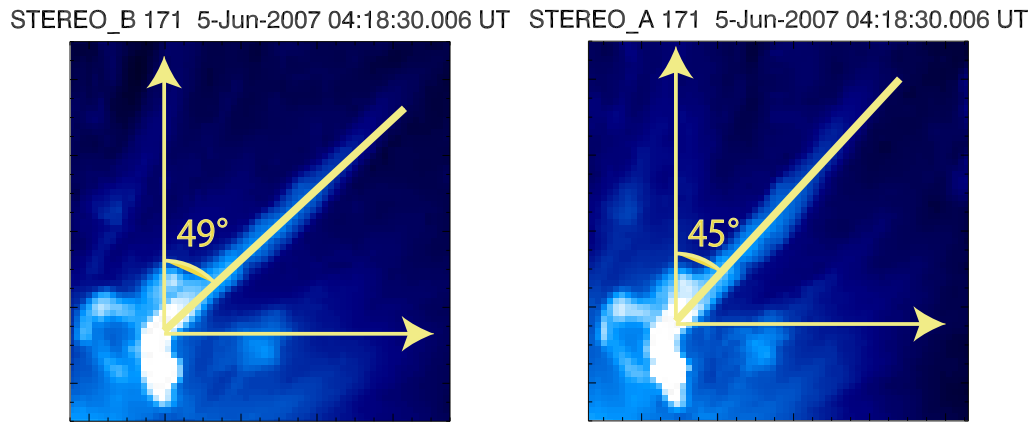}
 \caption{The left panel shows J1 observed from {\it STEREO-B}, and
the right panel shows J1 observed with {\it STEREO-A}. Yellow lines represent the
 observed inclination angle of J1 in each plane.\label{stereo_img}}
\end{figure}

\begin{figure}
\begin{center}
 \includegraphics[width=70mm]{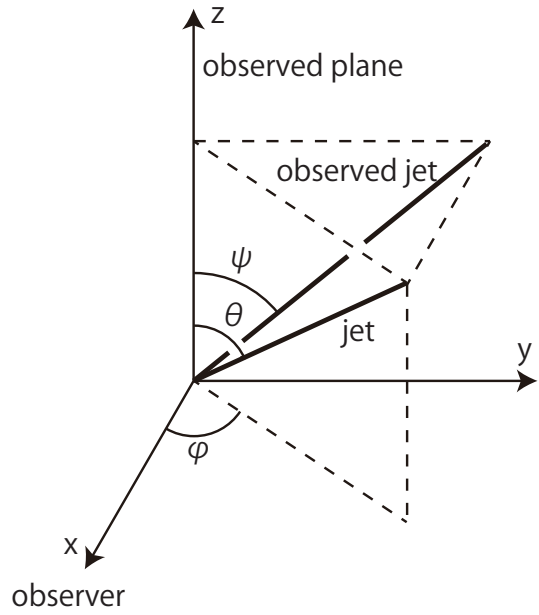}
 \caption{The polar coordinate system view of the jet. The $x$-axis is
toward the observer and $z$-axis is parallel to the direction of the solar north
 pole. The origin of this system is the footpoint of the jet.
 \label{inclination}}
\end{center}
\end{figure}

The three-dimensional structure can be obtained from the {\it STEREO-A} and {\it STEREO-B}
images taken from the different view angle.
The right (left) panel of figures \ref{stereo_img} shows J1 observed with
{\it STEREO-A} ({\it STEREO-B}).
Figure \ref{inclination} is a polar coordinate system view of the jet.
In this coordinate system, the $x$-axis is toward the observer and
the $z$-axis is parallel to the direction of the solar north pole.
The origin of this system is the footpoint of the jet.
The direction of the $x$-axis depends on the position of the observer
as in the case of {\it STEREO-A} and {\it STEREO-B}.
The origin and $z$-axis do not change for any observers.
The inclination and azimuth of the jet are $\theta$ and $\phi$,
respectively.
The observed jet in figure \ref{inclination} is the projection to the
$yz$-plane.
The apparent inclination of the observed jet is $\psi$ in this figure.
This angle is expressed using $\theta$ and $\phi$,
\begin{eqnarray}
 \label{eq1}
 \tan \psi = \tan \theta \sin \phi .
\end{eqnarray}
From the position of {\it STEREO-A} and {\it STEREO-B}, the azimuths are
 $\phi_{\rm{A}} = \phi_{\rm{E}} - \alpha$ and $\phi_{\rm{B}} = \phi_{\rm{E}} +
 \beta$ respectively, where $\phi_{\rm{E}}$
 is the azimuth for observers at the Earth.
$\alpha$ ($\beta$) is the angle between {\it STEREO-A} ({\it STEREO-B}) and the earth around
 the sun.
 The inclination $\theta$ is independent of the position of the considering observers.
 The apparent inclinations observed with {\it STEREO-A} and {\it STEREO-B} are
 $\psi_{\rm{A}} = 45^\circ$ and  $\psi_{\rm{B}} = 49^\circ$,
 respectively, and are overplotted in figure  \ref{stereo_img}.
 $\psi_{\rm{A}}$ and $\psi_{\rm{B}}$ are described using equation (\ref{eq1}),
\begin{eqnarray}
\frac{ \tan \psi_{\rm{A}}}{ \tan \psi_{\rm{B}}} =  \frac{ \sin (\phi_{\rm{E}} - \alpha )}{ \sin(\phi_{\rm{E}} + \beta )}. \label{tan_an}
\end{eqnarray}
 When {\it STEREO-A} and {\it STEREO-B} observed the jet, their locations are at
 $\alpha = 7.24^\circ$ and $\beta = 4.07^\circ$.
The azimuth and inclination are derived from equation (\ref{tan_an}) and
$\phi_{\rm{E}} = 56.3^\circ$ and $\theta = 52.9^\circ$, respectively.

  \subsection{Velocity}
\begin{figure}
\begin{center}
 \includegraphics[width=100mm]{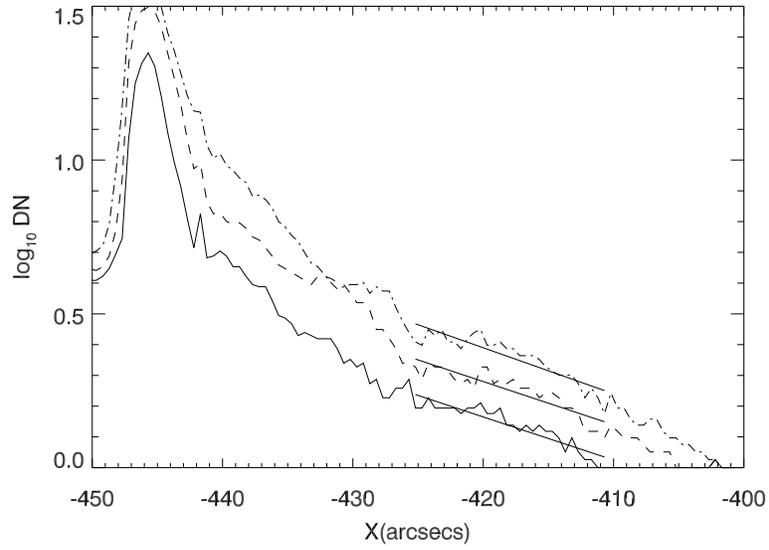}
 \caption{ Intensity distributions along J1 observed with
{\it TRACE} 171 \AA \  at the 04:16 UT in solid line, 04:17 UT in dash
line, and 04:18 UT in dash-dot line.
 \label{fig5}}
\end{center}
\end{figure}

Figure \ref{fig5} is the intensity distribution along J1 observed with
{\it TRACE} 171 \AA \ at the 04:16 UT in solid line, 04:17 UT in dashed line,
and 04:18 UT in dash-dotted line. 
The distributions of the intensity in each exposure time are well described by an exponential function, that is consistent with \cite{1996PASJ...48..123S,2001ApJ...550.1051S}.
The temporal offsets of the intensity profile have a speed of $155
\rm{km/s}$.  We interpret this as a projection velocity of the plasma
motion along J1 at the temperature of this filter range.

\begin{figure}
 \begin{center}
 \includegraphics[width=150mm]{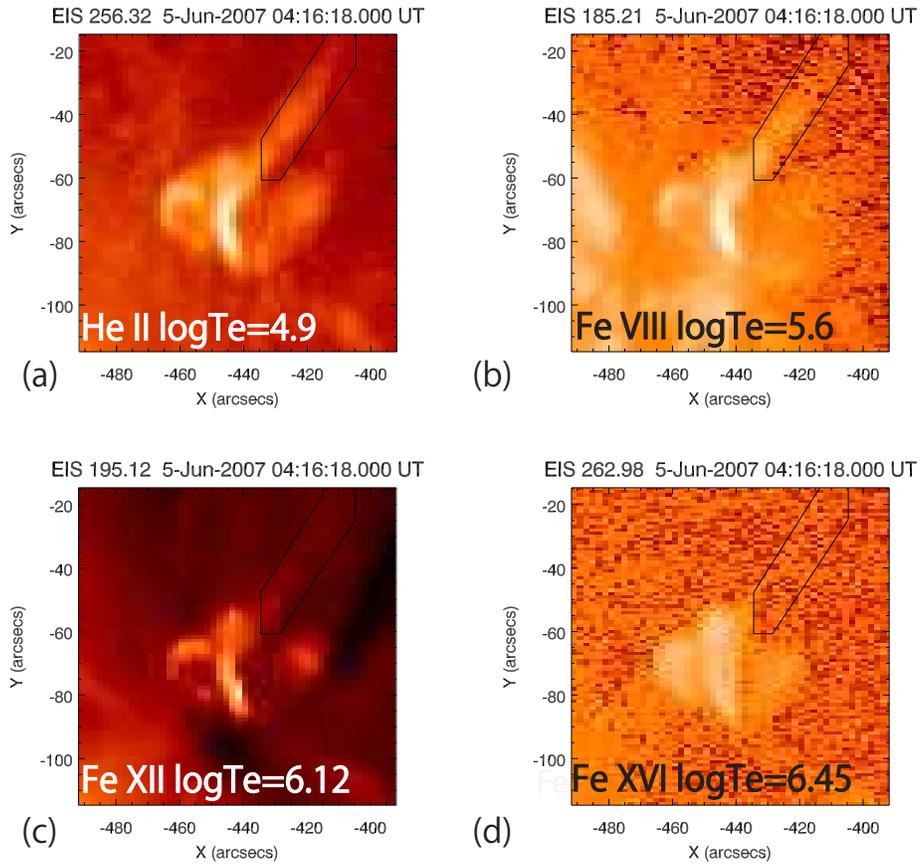}
 \caption{EIS intensity images of J1 observed with (a)
\ion{He}{2}, (b) \ion{Fe}{8}, (c) \ion{Fe}{12} and (d) \ion{Fe}{16},
  respectively.  
 The FOV of figures is a part of the EIS raster scan and corresponds
  to the {\it TRACE} and XRT images in figures \ref{fig1}.
 \label{int_map}}
\end{center}
\end{figure}

\begin{figure}
 \begin{center}
 \includegraphics[width=150mm]{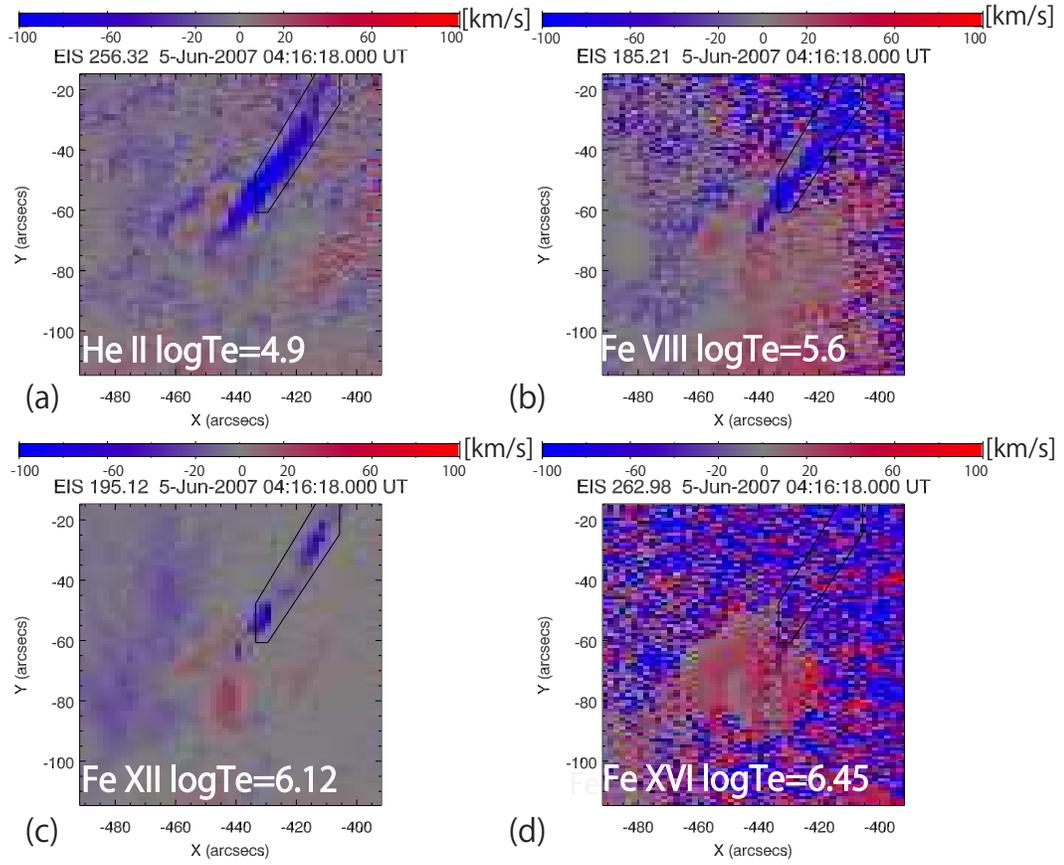}
 \caption{Doppler velocity images corresponding to the data
  set in figures \ref{int_map} (a)-(d), respectively.
 \label{vel_map}}
\end{center}
\end{figure}

EIS observed J1 with many emission lines.
The observed emission lines and their formation temperatures are summarized in table \ref{tb1}.
Figure \ref{int_map} shows the intensity images
 observed with (a) \ion{He}{2} 256 \AA \ ($\log_{10}T_e[\rm{K}] = 4.9$), (b) \ion{Fe}{8} 185 \AA \ ($\log_{10}T_e[\rm{K}] = 5.6$),
 (c) \ion{Fe}{12} 195 \AA  \ ($\log_{10}T_e[\rm{K}] = 6.12$) and (d)
 \ion{Fe}{16} 262 \AA \ ($\log_{10}T_e[\rm{K}] = 6.45$), respectively.
The field of view (hereafter, FOV) of figure \ref{int_map} is a part of the EIS
raster scan and corresponds to that of the {\it TRACE} and XRT images in figure \ref{fig1}.
The scan of figure \ref{int_map} started at 04:16 UT and ended at
04:21UT.
Figures \ref{int_map} (a) and (b) are the examples of the cool lines and
figures \ref{int_map} (c) and (d) are the examples of the hot lines.
The jet is clearly identified in the cool lines while it is not clear in
the hot lines.
The Doppler velocity images are shown in figure \ref{vel_map} (a)-(d).
Doppler velocities are obtained by fitting the
spectrum with a single Gaussian function and calculated from the displacement of
the center of the fitting Gaussian function using the eis\_auto\_fit in SolarSoftWare.
The blueshift of the jet is clearly shown in figures \ref{vel_map}
(a), (b) and (c) but is not clear in figure \ref{vel_map} (d) because of
the weak intensity and large noise of \ion{Fe}{16} 262 \AA. 

\begin{figure}
 \begin{center}
  \includegraphics[width=150mm]{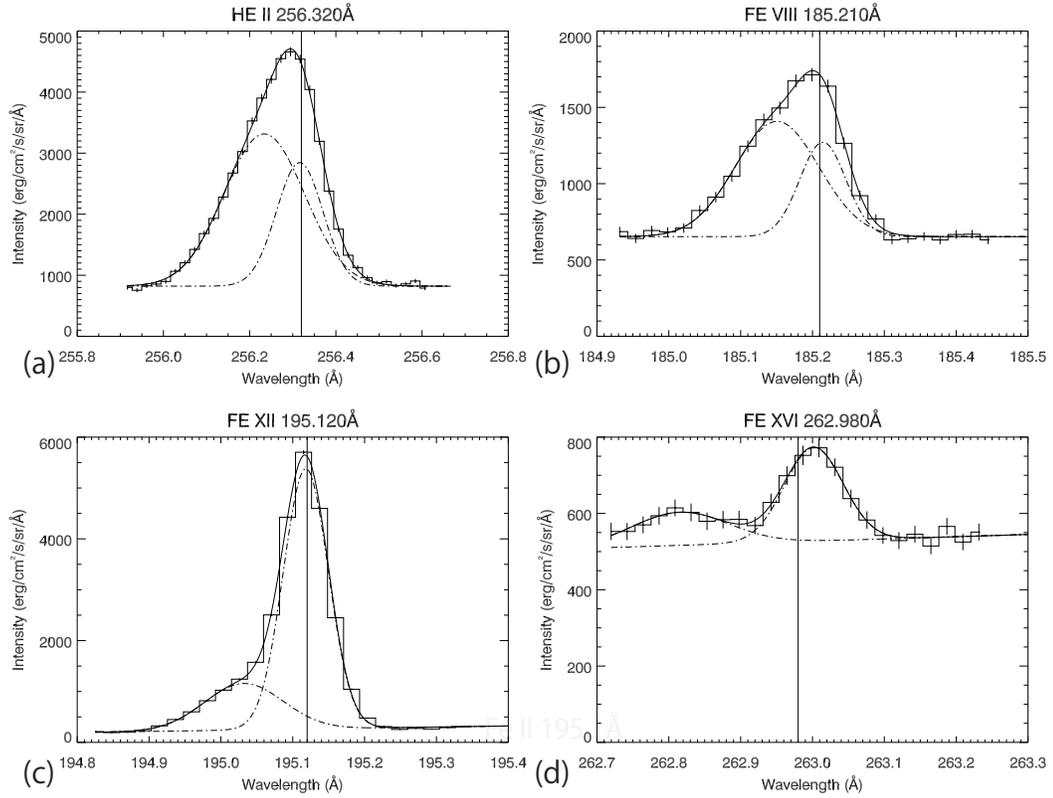}
 \caption{Line profiles averaged in the jet region.
Panels (a)-(d) correspond to the data set for figures
  \ref{int_map}(a)-(d), respectively. The histogram is the data with
  photon-noise error bars, solid line is the fitted profile and dash-dotted
  lines are its components. The vertical solid line indicates
 the line center of the emission line.
 \label{line}}
\end{center}
\end{figure}

We determine the jet region by a visual inspection shown by the black lines
in figures \ref{int_map} and \ref{vel_map}.
In this region, the intensity of \ion{He}{2} is bright.
\ion{He}{2}, \ion{Fe}{8} and \ion{Fe}{12} show the strong blueshift in this
region.
Figure \ref{line} shows the line profiles averaged in the jet region.
All lines show the strong blueshift even in \ion{Fe}{16}.
The blending effect of the neighboring \ion{Si}{10} 256.37 \AA \ 
($\log_{10}T_e[\rm{K}] = 6.2$) with the \ion{He}{2} blueshifted
component is weak because it can be estimated by using the ratio of the
blueshifted component to the rest component of \ion{Fe}{12}.
The effect of the hotter line \ion{Fe}{17} 262.8 \AA \ around the
blueshift component of \ion{Fe}{16} is neglectable because the expected
intensity from the CHIANTI atomic database
\citep{2006ApJS..162..261L,1997A&AS..125..149D} and \ion{Fe}{17} 254.9
\AA \ observed with EIS is much smaller than that of the blueshift component.

\begin{figure}
\begin{center}
 \includegraphics[width=150mm]{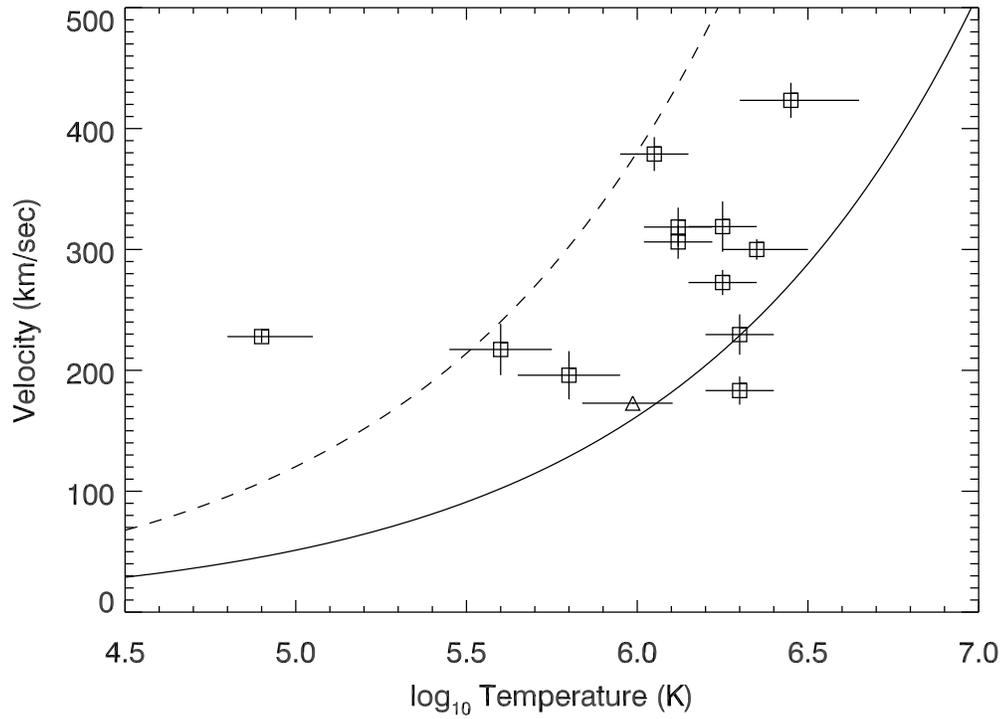}
  \caption{The relationship of the temperature and velocities observed
 with EIS and {\it TRACE}. The jet velocities are derived from Doppler shifts
 observed with EIS (square symbols) and projection velocity observed
 with {\it TRACE} (triangle symbol) using equations (\ref{eq3}) and
 (\ref{eq4}). Solid line shows the sound speed and 
 dashed line shows a theoretical upper limit to chromospheric
 evaporation velocities by \cite{1984ApJ...281L..79F}.
 \label{vel}}
 \end{center}
\end{figure}

The spectrum of each line (summarized in table 1) is averaged in the jet
region and fitted with a double Gaussian function to derive the
Doppler velocities of the blueshift component.
The true velocity of the jet is derived from the Doppler velocity and projection
velocity,
\begin{eqnarray}
 V_{\rm{jet}} & =& \frac{V_{\rm{Doppler}}}{\sin \theta \cos \phi_{\rm{E}}}  \label{eq3} \\
  & =& \frac{V_{\rm{projection}}}{\sqrt{1 - \sin ^2\theta \cos
   ^2\phi_{\rm{E}}}}  \label{eq4}, 
\end{eqnarray}
where $V_{\rm{jet}}$, $V_{\rm{Doppler}}$ and $V_{\rm{projection}}$ are the
jet velocity, Doppler velocity and projection velocity, respectively.
The azimuth $\phi_{\rm{E}}$ and inclination $\theta$ are derived by equation
(\ref{tan_an}).
The relationship of the temperature and
thus derived true jet velocity
is shown in figure \ref{vel}.
The temperature of the jet is defined
by the line formation temperature
(table \ref{tb1}) and
its error is estimated by
the FWHM of the contribution function based on the CHIANTI atomic database.
The error bar of the velocity indicates the error of
the double Gaussian fitting.
The velocity derived from the projection velocity of {\it TRACE} 171 \AA \
is indicated by the triangle symbol.
The temperature of this component
is defined by the temperature response of {\it TRACE} 171 \AA \ filter
\citep{1999SoPh..187..229H}.
The FWHM of the temperature response is shown as an error bar of the
triangle symbol in figure \ref{vel}.
The solid line in the figure \ref{vel} shows the sound speed and the dashed line
shows a theoretical upper limit to the chromospheric evaporation velocities
derived in \cite{1984ApJ...281L..79F}.

\subsection{Magnetic and Chromospheric Features}
\begin{figure}
  \begin{center}
 \includegraphics[width=150mm]{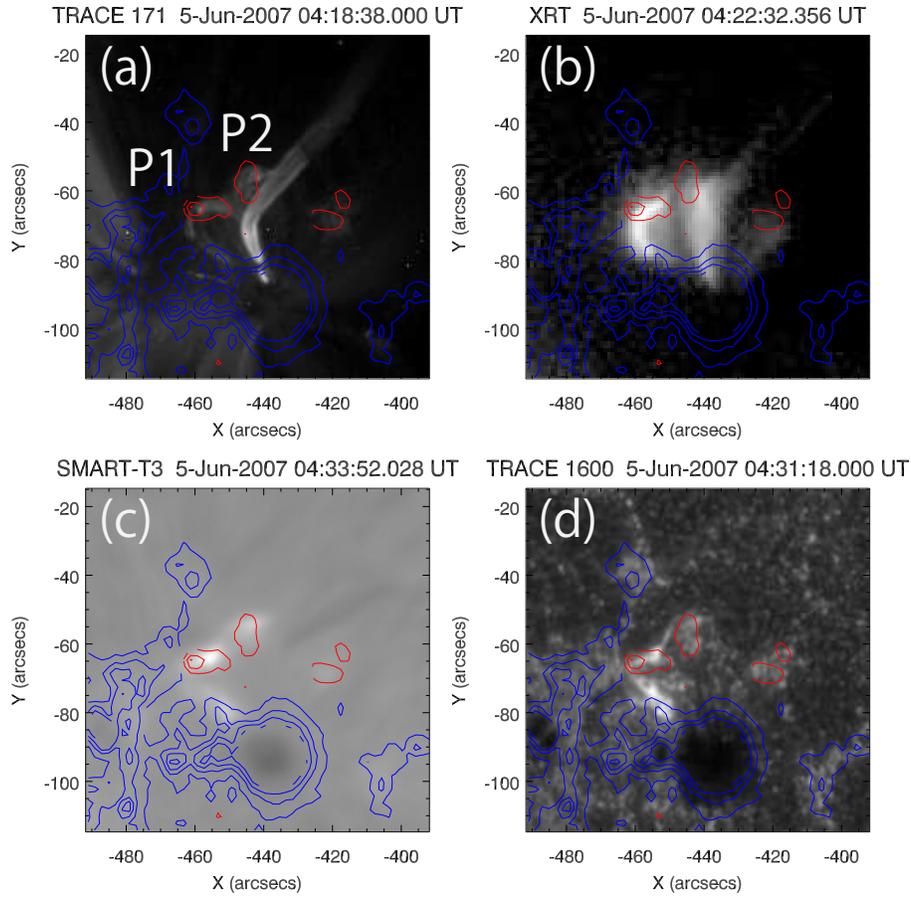}
 \caption{Red and blue contours represent the positive and negative
   line-of-sight magnetic field observed with {\it SOHO}/MDI respectively.
 Background images show (a) the {\it TRACE} 171 \AA \ intensity image at J1
   occurrence time,
(b) the X-ray intensity taken with the Ti\_poly filter of XRT, (c) the
   H$\alpha$ image taken with SMART-T3 and (d) the {\it TRACE} 1600 \AA \ image after the J2 occurrence time.
 \label{fig8}}
    \end{center}
\end{figure}

Figure \ref{fig8}(a) shows the magnetic field observed with {\it SOHO}/MDI.
The red and blue contours represent the positive and negative line-of-sight
magnetic field respectively, and the background shows the {\it TRACE} 171 \AA \ 
intensity at the J1 occurrence time.
It is found that J1 occurred near the large negative sunspot and the
positive polarities (P1 and P2).
By the overlay plot on the X-ray image in figure \ref{fig8}(b), we find
that the eastern (L2' in figure \ref{fig1}(h)) and western (L3) loops are
connecting P1 and P2 with the sunspot, respectively.
Figures \ref{fig8}(c) and (d) show the H$\alpha$ and
1600 \AA \ intensities at the J2 occurrence time (see also figures \ref{fig1}(e)
and (j)). In both panels, we see two ribbons each of which
corresponds to each magnetic polarity.
One of the ribbons is located at P1's position,
suggesting that they are footpoints of loop L2'. We will discuss the
geometrical magnetic structure driving the jets in section \ref{magfi}
based on these observations.

\section{Discussion}
\subsection{Geometrical Interpretation of Global Magnetic Structure
  \label{magfi}}
\begin{figure}
  \begin{center}
 \includegraphics[width=150mm]{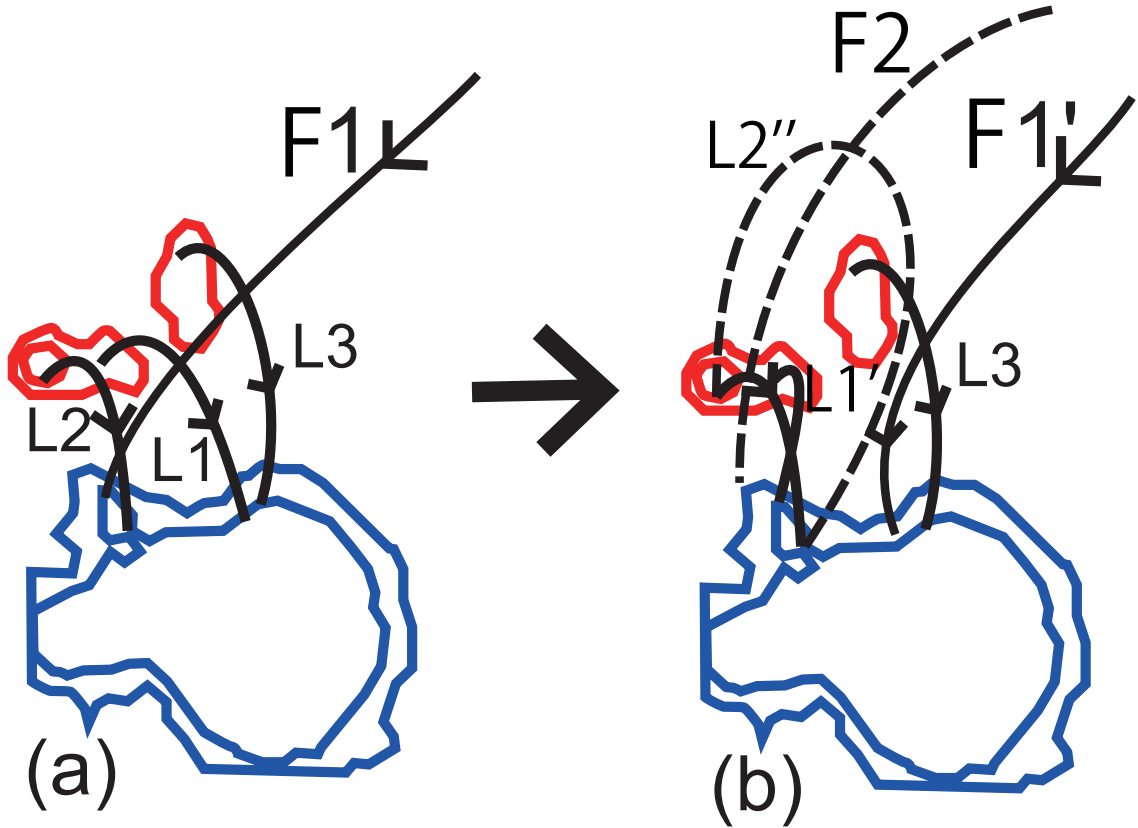}
 \caption{Schematic pictures of our interpretation on the geometrical magnetic structure.
 The red and blue lines are the positive and negative magnetic
field on the photosphere observed with MDI respectively, and black lines
represent the magnetic field.
The left (right) positive red polarity is P1 (P2).
   Each panel shows (a) before the onset of J1 and (b) around the time
   of J1.
 \label{fig9}}
  \end{center}
\end{figure}

We consider the global magnetic field structure (figure \ref{fig9}) consistent with the
observed jets, loops (figure \ref{fig1}), magnetic fields and flare
ribbons (figure \ref{fig8}).
The red and blue lines are the positive and negative magnetic
field on the photosphere observed with MDI respectively, and black lines
represent the interpreted magnetic field lines.
Before the onset of the jet (\ref{fig9}(a)), an open magnetic field F1
is connected with the sunspot.
There are also closed magnetic loops (L1, L2) between the positive magnetic
field P1 and the sunspot.
Another positive magnetic field P2 is also connected with the sunspot by a
closed loop (L3).
The first jet J1 is produced by the reconnection which occurred between
the closed loop L1 and open field F1 in figure \ref{fig9}(a).
This scenario is a kind of a standard jet \citep{1994xspy.conf...29S}.
As a result of the reconnection between these two fields, field lines F1'
and L1' are produced.
F1' is greatly bent because of L3 in figure 10(b).
F1' and L3 reconnect again and L3 becomes bright in figure 1g around y=80''.
L3 seems to be filled with the dense plasma caused by the
chromospheric evaporation.
L3 also shows redshift in figure 6c, which might be a shrinking motion of the reconnected field lines with plasma.
Flare ribbons in figure \ref{fig8}(c) and (d) around the footpoints of
the L1' and L2 are thought to be due to the another reconnection
between L1' and L2.
The second jet J2 is considered to be related to another open magnetic
field F2 described in figure \ref{fig9}(b) as a dashed line, which is
not seen in the observations.
Another possibility for J2 is a blowout eruption of sheared core field
around L2 \citep{2010ApJ...720..757M}.
In this interpretation, L2'' in figure \ref{fig9}(b) is the erupted loop.
This eruption is triggered by the reconnection that produced J1.
In this case, the first standard jet creates the second blowout jet.

\subsection{Magnetic and Thermal Acceleration
    \label{accele}}
EIS observed strong blueshifts in the multiple lines covering the wide temperature range as shown in figure \ref{vel}.
The profiles in figure \ref{line} show the blueshifts of the observed lines.
At the high temperature (\ion{Fe}{12} and \ion{Fe}{16}), the intensities
of the blueshift components are weaker than the rest one.
At the low temperature line (\ion{He}{2} and \ion{Fe}{8}), on the other
hand, the intensities of the blueshift components are stronger than the
rest one.
This implies that the density of the cool plasma in the jet is larger than
that of background cool plasma, which probably relates to the difference
of the acceleration of the cool and hot plasma though the reason
therefor is not revealed clearly in this study.

Finally, we discuss the difference of the acceleration of the cool and hot plasma.
We consider the two types of the acceleration, i.e. the magnetic acceleration by the magnetic reconnection and the thermal acceleration by the chromospheric evaporation.
The thermal accelerations like the chromospheric evaporation
\citep{2009ApJ...699..968M} and dimming flows
\citep{2007PASJ...59S.793I,2011ApJ...743...57I} show a dependence of the
velocity on the temperature.
If the jet is accelerated by the chromospheric evaporation, the velocity
of the jet increases with increasing temperature along with the sound speed.
Its theoretical upper limit is approximately $2.35C_s$ \citep{1984ApJ...281L..79F}.
The velocities of the hot plasma described in figure \ref{vel}
show a tendency to change along with the sound speed within this upper
limit, which is consistent with the chromospheric evaporation.
On the other hand, the velocity of \ion{He}{2} is much faster than this
upper limit.
This probably suggests that \ion{He}{2} is accelerated by the magnetic acceleration.
We interpret that the magnetic acceleration also occurs in the jet
along with the chromospheric evaporation.
In other words, hot plasma is accelerated by the chromospheric evaporation
while cool plasma is accelerated by the magnetic force.
This magnetic acceleration of the cool plasma is consistent with a blowout jet
\citep{2010ApJ...720..757M}.
In this jet, two types of accelerations, i.e. the magnetic and thermal accelerations occur at the same time. 
This result is consistent with the magnetic reconnection model.

\section{Summary}
We analyze the fine structure of a solar EUV jet based on the magnetic
reconnection model. Multi-wavelength spectral observations with EIS
allow us to know the Doppler velocities at the wide temperature range.
We find that this jet is accelerated by the thermal acceleration and magnetic acceleration at the same time.
Our results are consistent with the magnetic reconnection model.
To confirm our conclusion, we will perform the numerical simulation and
reproduce the jet under the similar condition of this observation in the future work.

\vspace{1cm}

{\it Hinode} is a Japanese mission developed and launched by
ISAS/JAXA, collaborating with NAOJ as a domestic partner,
NASA and STFC (UK) as international partners. Scientific operation
 of the {\it Hinode} mission is conducted by the {\it Hinode}
science team organized at ISAS/JAXA. This team mainly consists
 of scientists from institutes in the partner countries. Support
 for the postlaunch operation is provided by JAXA and
NAOJ (Japan), STFC (UK), NASA, ESA, and NSC (Norway).
We are deeply grateful for the proofreading/editing assistance from the GCOE
program.

\bibliography{ref1}

\clearpage
\begin{table}
\caption{The observed emission lines and their formation temperatures.\label{tb1}}
\begin{tabular}{cccc}
\tableline\tableline
 Ion & Wavelength $\lambda (\rm{\AA})$ & $ \log_{10}T_e[\rm{K}] $  \\
\tableline
 \ion{He}{2} &256.32 &4.90  \\
 \ion{Fe}{8} &185.21 &5.60 \\
 \ion{Si}{7} &275.35 &5.80  \\
 \ion{Fe}{10} &184.54 &6.05 \\
 \ion{Fe}{12} &186.88 &6.12  \\
 \ion{Fe}{12} &195.12 &6.12  \\
 \ion{Fe}{13} &202.04 &6.25  \\
 \ion{Fe}{13} &203.83 &6.25  \\
 \ion{Fe}{14} &274.20 &6.30  \\
 \ion{Fe}{14} &264.78 &6.30  \\
 \ion{Fe}{15} &284.16 &6.35  \\
 \ion{Fe}{16} &262.98 &6.45  \\

\tableline
\end{tabular}
\end{table}

\end{document}